\documentclass[review]{elsarticle}
\usepackage[margin=1in]{geometry}
\usepackage{lineno,hyperref}
\usepackage{dirtytalk}
\modulolinenumbers[5]
\def\bsq#1{
\lq{#1}\rq}

\journal{Journal of \LaTeX\ Templates}

\begin{document}

\begin{frontmatter}

\title{Big Data Analytics and AI in Mental Healthcare}

\author{Ariel Rosenfeld$^{*}$}
\address{Bar-Ilan University, Israel}

\author{David Benrimoh$^{*}$}
\address{McGill University, Canada. Aifred Health}

\author{Caitrin Armstrong, Nykan Mirchi, Timothe Langlois-Therrien, Colleen Rollins, Myriam Tanguay-Sela, Joseph Mehltretter, Robert Fratila, Sonia Israel, Emily Snook, Kelly Perlman}
\address{Aifred Health}

\author{Akiva Kleinerman}
\address{Bar-Ilan University, Israel}

\author{Bechara Saab, Mark Thoburn}
\address{Mobio Interactive}
 
\author{Cheryl Gabbay}
\address{McGill University, Canada}

\author{Amit Yaniv-Rosenfeld}
\address{Tel-Aviv University, Israel}

\begin{abstract}
Mental health conditions cause a great deal of distress or impairment; depression alone will affect 11\% of the world's population. The application of Artificial Intelligence (AI) and big-data technologies to mental health has great potential for personalizing treatment selection, prognosticating, monitoring for relapse, detecting and helping to prevent mental health conditions before they reach clinical-level symptomatology, and even delivering some treatments. However, unlike similar applications in other fields of medicine, there are several unique challenges in mental health applications which currently pose barriers towards the implementation of these technologies. Specifically, there are very few widely used or validated biomarkers in mental health, leading to a heavy reliance on patient and clinician derived questionnaire data as well as interpretation of new signals such as digital phenotyping. In addition, diagnosis also lacks the same objective ‘gold standard’ as in other conditions such as oncology, where clinicians and researchers can often rely on pathological analysis for confirmation of diagnosis. In this chapter we discuss the major opportunities, limitations  and techniques used for improving mental healthcare through AI and big-data. We explore both the computational, clinical and ethical considerations and best practices as well as lay out the major researcher directions for the near future.
\end{abstract}

 \end{frontmatter}


\section{Introduction}
 The conceptualization, diagnosis, treatment and prevention of mental disorders is limited by existing options for collecting, organizing and analyzing information. Big data and machine learning/artificial intelligence (ML/AI) can be applied to the development of tools that could help patients, providers and systems overcome these limitations. As many as 1 in 5 people  experience mental illness. Mental disorders affect individuals' abilities to function, engage meaningfully in daily activities and maintain relationships. They  cause significant suffering to individuals and their families and are a significant source of  socio-economic burden \cite{AMI}. Many mental disorders are also risk factors for suicide which occurs at an alarming rate globally \cite{WHO}. These are disorders that often strike young and otherwise healthy people, a socially and economically critical segment of the population. As such, improving the detection, treatment, and monitoring of mental illness is crucial.
 
 Designing tools for practical use-cases in mental healthcare requires a deep understanding of psychiatric illness, the current mental healthcare system, and medical ethics. We begin with an introduction of mental illness and healthcare, and proceed to discussing their complexities from a clinical and data-driven perspective, before discussing specific use cases and applications of big data and machine learning approaches. To the reader from engineering and computer science: perhaps the most important conclusion from this chapter is that close collaboration with domain experts and clinicians will be required invariably in order to successfully build safe, effective, and useful mental healthcare applications.
 
Mental illnesses are a group of diverse conditions with varying severity, complexity, and duration. In considering deviations from normal thought, feeling and behavior, characteristic of mental illness, it is critical to recognize the extent to which they lead to functional impairment. To be classified as a disorder, a set of symptoms must cause significant suffering or interference with daily functions or life goals \cite{DSM}. This means that the treatment of mental illnesses has largely the same objective as other branches of medicine: the alleviation of suffering, the improvement of function and quality of life, and the reduction of morbidity (the incidence of new diseases or impairments) and mortality (or the rate of death, primarily from suicide or reduced life expectancy because of impaired self-care). These then are the objectives of the clinical professionals who treat patients with mental illness. Family doctors are the primary providers of mental healthcare  \cite{fleury2012general} and are accompanied by many other healthcare workers and specialists such as  psychiatrists, psychologists, nurses, social workers, case managers, occupational therapists, pharmacists and counselors. Approaches to measurement and treatment within mental healthcare differ significantly from other areas of healthcare.  An example of an approach that is widely used in medicine but not useful in mental healthcare is diagnosis confirmation through pathological examination (i.e. via examination of patient tissues). We will highlight other such differences throughout this chapter.
 
There is a diversity of mental illnesses with varying  clinical presentations, time-courses, and causes. Autism, schizophrenia, work-related burnout, dementia, attention-deficit hyperactivity disorder (ADHD), eating disorders, and addictions, along with many others, are all included under the banner of mental illnesses. As such, just as there are many different versions of cancer whose causes, genetics, disease courses, and prognoses are very different from each other (and which can have different severities and types even within the same disease), mental illnesses present a kaleidoscope of different conditions. An in-depth discussion of these different disorders and their presentations is beyond the scope of this chapter. There is also a considerable amount of individual variation within disorders. For example, people with Major Depressive Disorder (MDD), otherwise known as depression, can present with sadness, guilt, feelings of inadequacy, lack of sleep and a profound lack of energy. Other patients with depression can present with lethargy, overeating, oversleeping, poor concentration, and thoughts of suicide \cite{DSM}. Mild cases of depression can often respond well to exercise or psychotherapy, whereas more severe cases may  only respond when medication or even electroconvulsive therapy is included in their treatment plan \cite[Section~1.10]{NICE}. Some patients present with recurrent experiences of  depression while  others with present with a singular depressive episode. Some depressive episodes are correlated with the experience of a triggering or traumatic event  such as the death of a loved one or a conflict at work while other depressive episodes do not seem to be correlated with any particular life event and begin seemingly \say{out-of-the-blue} \cite{malki2014endogenous}. The possible causes for these different presentations, and their effect on machine-learning and big-data approaches to improving mental healthcare, will be further discussed below.  Another important concept is comorbidity, or the existence in a single patient of more than one disorder. For example, a patient can present with both depression and ADHD. The incidence of comorbidity further complicates an already complex problem. Firstly, there is the challenge of diagnosing disorders with often overlapping symptomatology. Then there is the question of whether causality can be inferred or should be investigated. For example, depression can manifest as a reaction to an individual's difficulty coping with a pre-existing condition or it could be an entirely separate disease process. It is highly unlikely that one application or machine-learning model will be able to, given the diversity in mental health and the current state of technology, provide a tool that would be useful across  all of these conditions. Favoring development aimed at specific use-cases, informed by an understanding of a particular disorder, will likely result in more fruitful efforts.
 
How are these diverse conditions treated? The most important thing to understand is that there is no singular treatment that is effective in all presentations of a disorder or appropriate for every individual. Treatment involves more than just diagnosis, drug prescription and/or psychotherapy. Perhaps more than in other branches of medicine, it is critical in mental health provision to dedicate the time to form an alliance or partnership with the patient and to understand the patient's life situation, goals, social network, belief system, and personal and community resources. Only when one understands a patient in this way and has built a \textit{therapeutic alliance} \cite{ardito2011therapeutic}, a trusting professional relationship, can appropriate treatments or interventions be implemented successfully. As such, over time, big data must evolve to capture these elements of patient care, and machine learning analyses must incorporate them as critical elements to model.
 
Because of the need to understand and support the patient as just described, treating severe mental illness is often not practical unless a \textit{team} of professionals is involved \cite{bond2015critical}. For example, a psychiatrist may manage medications while a nurse  monitors medication blood levels and  side effects. Furthermore, a social worker may help the patient manage finances while a psychologist works with the patient in therapy, and a family doctor manages non-psychiatric conditions such as diabetes. As such, any clinically-focused applications or big data collection initiatives should consider these many sources of information and the multiple agents that are involved in the clinical decision-making pathway. 
Furthermore, both the patient and (sometimes) their family, are active participants in planning, choosing, arranging and prioritising the care they receive.  An understanding and assessment of individual choices and preferences and how these interact with the clinical team's decision-making process is another important piece of the puzzle.
 
The journey experienced by a patient- what might be termed as the \say{user story} in other application contexts- may vary wildly as a function of individual resources and of the specific disorder the patient has. Let us take as an example the journey of a patient who develops schizophrenia (referred to as \say{she}), and contrast this with one who develops depression (referred to as \say{he})- both to highlight differences and similarities between the two and to demonstrate how the mental healthcare system interacts with patients, as both are relevant to how and where big-data and machine-learning-powered methods may be integrated into practice. The first patient is a university student, with an unremarkable family history and a stable home environment. At the age of 22, she begins experiencing strange things- whispers that come out of nowhere, fleeting thoughts that her friends wish her harm, and difficulty focusing on coursework. She begins to isolate herself, and her friends start growing concerned after she stops coming to class. After six months of this progressively worsening situation, she is brought to the hospital, by the police, because she was yelling on the campus square about an imminent attack by some undefined group. She is brought to an emergency room, sedated, and in the morning she is assessed by a psychiatrist and started on an antipsychotic medication; no firm diagnosis is given, only the acknowledgement that she had a psychotic episode and that no drugs seemed to be involved. Once the patient has returned, somewhat, to their usual self, they are referred to an outpatient program for further evaluation. In this program they are seen by another psychiatrist, a nurse, and a case manager- all of whom ask many questions in long interviews. Still no firm diagnosis is made, as it is too early to say anything definitive. The patient is recommended to continue taking her antipsychotic medication, but the she refuses because she is afraid of weight gain (a common side-effect). The patient eventually has another psychotic episode four months later, spends two months on a locked inpatient unit recovering, and is diagnosed formally with schizophrenia. Because of concerns that the patient will stop taking the medication again, she is recommended to take- and agree to- a long-acting injectable form of the drug, which helps her stabilize and return to her studies. In this example, the patient is presenting to the emergency room, the outpatient program, and the inpatient program; she initially does not take her medication or adhere to follow-up, but eventually improve with the recommended treatment.

The second patient, who suffers from depression may face a very different experience. At the age of 30, his performance at work begins to lag, and his boss suggests that they go and speak to the company counselor. After a brief conversation, the counselor suggests that the patient is experiencing more than burnout and suggests they visit their doctor. The family doctor diagnoses the patient with depression, and notes that the abuse the patient suffered in his childhood serves as a significant risk factor for the disorder. The patient agrees to try taking antidepressants, but stops taking the drug, with their doctor's agreement, after four weeks because it does not seem to be working and causes sexual side effects. The patient then tries psychotherapy, but because of the long wait before the start of therapy (because of a long waiting-list) and a poor therapeutic alliance with the therapist, the patient worsens. He return to his family doctor and decide to try one more medication and this one begins to work; within three weeks the patient is feeling better and in two months they are back at work. Here, the patient never even saw a psychiatrist and was treated only by a family doctor and a psychologist; they had to go through several treatments, even though they adhered to each one.

The two patients discussed above had very different journeys and challenges. As a result,  any computational model or application developed to support them would need to be targeted at the appropriate part(s) of their prospective journeys. 
 
The reminder of this chapter is structured as follows: In Section \ref{sec:complex}, we  examine further details of what makes mental healthcare complex from clinical and data-driven perspectives. In Section \ref{sec:elements}, we  deconstruct the patient journey down to the most common elements or steps and discuss the challenges, opportunities and use-cases for each of these, as well as important ethical considerations.  Last, in Section \ref{sec:conclusions} we discuss and summarize the content of this chapter.

\section{What Makes Mental Healthcare Complex?}\label{sec:complex}

In this section we will discuss some of the challenges inherent in applying machine-learning to big-data approaches in mental health. A fundamental challenge to treating mental health problems is the lack of a \textit{mechanistic model} for essentially any psychiatric disorder \cite{kendler2008explanatory,huys2016computational}. Specifically, in contrast to some other branches of medicine like cardiology, we do not understand the mechanisms that lead to and sustain states such as depression or psychosis. While a wealth of results from psychological, genetic, environmental, metabolomic, epidemiological and neuroimaging approaches have advanced our understanding of the causes of psychopathology, it remains the case that we do not have a clear model that describes the development of a given mental illness, unlike the way we understand how plaque builds up in arteries and leads to a heart attack. This is not due to a lack of effort or data; rather it can be attributed to the extraordinary complexity of understanding the human brain as it develops, processes information, and interacts with the continually-changing environment which we collectively shape through high-level processes such as culture and social hierarchies. In addition, while many studies have evaluated, for instance, the link between genetic markers and the etiology of schizophrenia \cite{marshall2017contribution} or the link between childhood abuse and the development of depression \cite{infurna2016associations}, it remains a challenge to integrate results across levels of investigation, from neurotransmitters and variations in genes, to brain systems, to psychological states or behaviours, and to the societal and cultural systems in which individuals are embedded. To express the problem in terms familiar to any computer or data scientist, we are attempting to interpret the functioning of the hidden layers of an extraordinarily complex neural network whose inputs and outputs are not easy to quantify and are subject to a lot of noise; in some respects, this is similar to the classic credit assignment problem \cite{minsky1961steps} where determining how the success or failure of a system is due to the various contributions of the system's components. 

Different branches of science (e.g. psychology, neuroscience, and artificial intelligence) make different theoretical assumptions about the nature of mental processes and take different stances on the \textit{mind-body problem} \cite[Section~1]{sep-dualism}. 
It appears that so far most data and evidence collected has been of insufficient quantity and/or quality, or has been of the wrong kind, to afford the construction of mechanistic models.            
This brings us to two practical challenges raised by the lack of mechanistic models: Choosing which data to record and how it should be represented in order to provide useful insights. Specifically, the lack of mechanistic models often leads to the analyses of large datasets in a ‘blind' manner- that is, by training simple classifiers or related data-driven algorithms in an attempt to find some model that explains the data. While recent deep learning advances \cite{lecun2015deep} can capture the implicit relationships between features in these models, they do not provide one with the desired mechanistic model. Furthermore, in most cases, these models turn out to be extremely difficult to interpret beyond a simple understanding of which features were most important in the model \cite{castelvecchi2016can}. That is, the complex moderation and mediation interactions these models discover are not always accessible to human researchers and cannot readily be translated into practical insights. 

Without a mechanistic model to test, we must rely on generating models from data. However, different datasets may produce different best-fit models, even when these datasets are large and the outcome being predicted is similar. A model that is significant in one dataset may or may not advance our understanding of the underlying disease, unless it replicates in other datasets and coheres with existing findings in the literature. Given the concern that bias engendered by training on non-representative datasets could creep into clinical applications, ensuring that models are generalizable is essential, which again, is very challenging. 
 
Let us turn to the challenge of measurement, or more precisely, knowing what to measure in a cost-effective manner. This is critical because having \say{big data} is only useful if the dataset actually contains informative variables. As discussed above, for our purposes, the type of data one expects to be informative depends on underlying theoretical assumptions. In addition, because there are so many possible types of data to collect, and because of the often high cost of data collection, it will not be feasible in most cases to expect to overcome this challenge by simply ‘collecting everything'. In making decisions about the type of data to collect, there are also practical considerations. For example, when designing big datasets, should the budget be spent on a collection of extensive neuroimaging data, which is difficult to apply in clinical settings, or on a set of very simple measures that can be more easily collected and applicable within the clinical setting but that may lack in explanatory power? 
Furthermore, there is also the challenge of operationalizing the constructs to be measured. Should we attempt to work with written case records via natural language processing- and if so, how do we deal with the differing terminologies, writing styles, and the fact that, often, two healthcare professionals with the same training will disagree on the diagnosis or treatment plan for the same patient \cite{aboraya2006reliability}? How do we combine existing datasets in a valid manner, understanding that they were all collected for different purposes and often using different tools and questionnaires? These are among some of the questions that must be addressed.

Unlike most other disorder types, mental illnesses diagnoses are based  entirely on their phenomenology- that is, on descriptions of symptom clusters and patient presentations amalgamated over time by the psychiatric and psychological community \cite{hyman2011diagnosing}. Sometimes diagnostic categories can seem arbitrary or overly rigid. For example, while anxiety symptoms are very common in depression, anxiety is not one of nine symptoms included in the official diagnosis of depression \cite{DSM} even though anxiety seems to be both neurobiologically related to depression \cite{tiller2013depression} and an important predictor of whether or not people will respond to antidepressant treatment \cite{saveanu2015international}. In order to be of clinical use, diagnostic tools based on big data and machine-learning should be validated against these admittedly imperfect diagnostic criteria. One caveat is that this need for validation against diagnostic criteria  may lead to a situation in which a tool that, for example, tracks user anxiety and sleep might be more reflective of the putative underlying neurobiology of depression, but may or may not have acceptable performance when detecting depression according to the formal criteria. This would depend on how well anxiety correlates with the other official symptoms. 

In terms of outcome measurement, in mental health this is usually accomplished by means of validated questionnaires such as the Quick Inventory of Depressive Symptomatology (QIDS) \cite{rush200316} or the Patient Health Questionnaire (PHQ-9) \cite{kroenke2001phq}. These are also often used for screening, determination of illness severity, and in some cases diagnosis. These questionnaires are themselves imperfect, with their accuracy varying depending on their length, whether they are patient-self report or clinician-rated, and in the latter case the level of training of the clinician. In addition, they are often validated by coherence with older questionnaires that measure the same construct. Other, \bsq{harder} outcomes can also be collected, such as employment,  service utilisation costs, and suicide. This, however, may be subject to strong correlations as a patient may or may not find employment because of prevailing economic conditions or other factors. This problem can be worsened because of the prevalence of comorbidity and the intersection of mental illness with social, cultural, and economic realities \cite{stewart2015socioeconomic} making it very difficult to understand the trajectories of many of the more severe patients in the mental healthcare system.  It is also important to note that although the patient is the expert on his or her own experience, the information that the patient can provide is limited especially in cases where executive functions such as attention, memory and others are impaired as is often the case in mental illness \cite{musso2014investigation}. 
These difficulties are compounded by several factors. Firstly, many patients see multiple providers \cite{rubin2012perspective} and can have chaotic or irregular contact with the health system. Furthermore, it can be difficult to accurately measure a patient's social situation both because of the challenge of operationalizing a patient's social network  and the limitations of available technologies. For example, even with the advent of social media monitoring, it is not clear whether these techniques capture important social connections \cite{santini2015association}. Furthermore, missing data is ubiquitous and can introduce bias into the dataset.
 
Given  the advent of big data (large volumes of heterogeneous variables) and improvements in processing power, machine-learning (see \cite{shalev2014understanding}) presents itself as a promising avenue to offer solutions to some of the aforementioned complexities inherent to treating mental health problems \cite{iniesta2016machine,passos2016big}. However, the promise comes in hand with challenges of applying big data analytics to mental health problems. Though the heterogeneity of clinical, sociodemographic, neuroimaging, genomic, immune, and other measures is advantageous from a research standpoint, it is challenging to deal with the impact of high dimensionality, especially when the number of features exceeds the number of subjects. On the other hand, much of the data collected may not be informative towards making a diagnosis or predicting treatment response, drastically reducing the number of features that are actually needed. The challenge is to separate the useful from the less useful features. It is equally important to consider the generalizability of machine-learning models  outside the training sample. Insufficient sample size and underrepresentation of minority groups also make it difficult to interpret machine-learning models for some populations.  An additional important challenge is that many machine-learning models cannot be easily understood by humans, commonly functioning as \say{black boxes}. 

\section{Opportunities and Limitations for AI and Big Data in Mental Health}\label{sec:elements}

Understanding the complexities inherent in the conceptualization, diagnosis and treatment of mental illness as well as how these complexities impact on the use of machine learning and big data in this space helps contextualize a discussion of potential and current use-cases for AI and Big Data in mental health. In the following section, we have grouped these use cases to reflect a patient's clinical course. 

The first step in this trajectory is \textit{diagnosis}. Mental illnesses are best treated if diagnosed early \cite{bird2010early}, and many cases of mental illness could even be prevented with the right interventions (including societal and population-level interventions) \cite{WHO}. Here, big data and machine-learning can help us better understand which people in the general population are at risk for developing mental illness, helping us deliver interventions that could save lives and costs by preventing the illness from fully manifesting or intensifying. On a population level, the use of machine-learning could help us identify population trends and variables that could be targeted by social programs to reduce the incidence of mental illness. 

Once a person is diagnosed or considered to be at-risk, they often want to know what their chances of recovery are, and many clinicians want to know about their risk of suicide or violence; this understanding of the likely course a patient's illness will take is called the \textit{prognosis}. Here,  predictive tools could help patients better understand their illness and help their families plan for different clinical courses. Clinicians do not perform well when trying to predict which patients are at risk of suicide, and so better tools for this have the potential to fill a clinical gap and save lives. 

Simply knowing a patient's prognosis does not necessarily lead to being able to help them manage their illness. For this, an optimal patient management strategy must be selected. This is where \textit{treatment selection} tools come in; these are tools aimed at using patient information to select the optimal treatment or intervention, from a range of psychiatric interventions that often do not separate by efficacy at the group level. 

Once a treatment is selected, one must choose how to \textit{deliver treatment}, and here, AI can help by providing virtual therapists or personalizing patient experiences in digital therapeutics applications. While these are unlikely to replace medications and traditional psychotherapy, they may prove to be a powerful ally in augmenting traditional therapies, improving access, and acting as low-intensity interventions that can be delivered in a preventative manner, before a patient requires more skilled or advanced care. 

Next, the \textit{monitoring} of patients' condition and symptoms is needed in order for clinicians to get a deeper understanding of the patient's illness. Typically, non-hospitalized patients do not see their clinician  on a very frequent basis and therefore the treatment is mostly based on the patients' (sometimes biased) report and condition during the appointments. Automated or semi-automated monitoring can mitigate this limitation. 

Finally, it is crucial to discuss the \textit{ethical considerations}: each of these use-cases must be with a respect for patient welfare, dignity, rights and current medical ethics, while endeavouring to address the more novel ethical considerations that come from the use of AI.

\subsection{Diagnosis}

Diagnosis is an important initial step in the treatment of mental illness and relies heavily on \textit{nosology}, the study of the categorization and explanation of disease. Three main nosological systems can be used \cite{kendler2008introduction}: 

\begin{enumerate}
    \item an etiological system (i.e. defined based on the cause(s) of a disease)
    \item a pathophysiological or mechanistic system (e.g. diabetes type 1 is defined through the absence of insulin production)
    \item  a symptom-based system (i.e., defined by clusters of symptoms)
\end{enumerate}

Psychiatry previously relied on an etiological nosological system centered around psychoanalytic theories. However, as discussed before, the field of psychiatry have moved to adopt the current \say{atheoretical} symptom-based model, as present in the Diagnostic and Statistical Manual of Mental Disorders \cite{DSM} from its $3^{rd}$ edition onwards. This radical paradigm shift ushered in issues of validity. The primary issue was and remains the fact that diagnosis in a symptom-based nosological model cannot be incorrect. That is, \textit{there is no other test or gold-standard diagnostic procedure; if the symptoms are present and meet criteria, the diagnosis is valid even if there may be reasons to doubt this conclusion}. Furthermore,  heterogeneity in symptom presentation and high levels of comorbidity  obscure the boundaries of the disease categories leading to high false positive and false negative rates \cite{klinkman1998false}. In response to these limitations, researchers have begun to  identify biomarkers and underlying neurobiological mechanisms to guide mental health diagnosis and treatment, though this work is still in the exploratory phase \cite{kendler2008explanatory}. 

Considering the difficulties facing psychiatric diagnoses, machine-learning and big data offer a unique opportunity to improve our understanding of and ability to diagnose diseases. We have identified three areas that could benefit from the right application of machine-learning and big data; namely, improved data collection, a better understanding of symptom clusters, and a redefinition of diseases with respect to function and quality of life. 
  
The popularization of AI, big-data approaches, and data collection technology development is encouraging the collection and analysis of unprecedented amounts of data, from a wider range of sources than ever before. Examples of these new data sources and analysis efforts in our context include:
\begin{itemize}
    \item The spread of electronic medical records, allowing for more access to healthcare data.
    \item  Social media which offers the opportunity to mine data to both inform diagnosis and examine the impact of disease. Internet users with various mental illnesses can be characterized by their social media use, text generation, and other online behaviours (e.g., \cite{birnbaum2017collaborative,de2013predicting,saha2016framework}). Social media analysis can also serve to inform our understanding of the sometimes-difficult changes in self-perception and identity that can occur after a diagnosis (see \cite{conway2016social}).
    \item Passive sensing of movement, location, social media, calling, and text message use through mobile phones \cite{wang2016crosscheck} (e.g., the Beiwe platform\footnote{\url{http://wiki.beiwe.org}}). This data may help us gain a deep understanding of the patterns and changes in patterns associated with mental illness, suicidality, and response to treatment.  
    \item Ambulatory assessment, which includes wearable sensor technologies that collect momentary data that do not depend on self-reports, context sensors (e.g. noise level, pollution), and biobehavioural sensors (which measure physical activity, sleep quality, blood pressure, alcohol intoxication and more) (Trull and Ebner-Priemer. 2013). Much like passive sensing from mobile phones, this kind of highly personalized data offers insights into realms of human behaviour and functioning, such as sleep and movement, that are directly relevant to diagnosis. 
\end{itemize}

Moreover, by accumulating all this information on patients, symptoms like \say{mood} or \say{appetite} that often mean different things to different patients could be more accurately assessed within a patient-specific model trained on one patient's past data to predict outcomes for them. 

AI may facilitate the deconstruction of clinical labels into more biologically-grounded transdiagnostic features. As we mentioned, the research community is struggling to understand underlying mechanisms specific to the currently labeled mental disorders. This is most likely explained by a many-to-many relationship between neurobiological processes and syndromes, or even symptoms. In other words, a neurobiological alteration could potentially give rise to multiple symptoms, and many symptoms could be resulting from more than one underlying process. Machine learning offers the chance to predict the different dimensions of symptomatology which a patient might experience. Meaningful clusters can then be found in this multi-dimensional landscape, and these might be stable across patients, allowing for new symptom clusters to be identified. For example, clustering techniques have been able to show important clustering between brain regions and some transdiagnostic features or dimensions of mental disorders, like mood, psychosis, disruptive behavior, or anhedonia \cite{xia2018linked,grisanzio2018transdiagnostic}. In these applications agglomerative hierarchical clustering techniques \cite{day1984efficient} were used to devise clusters of symptom scores, showcasing the power of machine-learning for psychiatry. This is in line with recent theoretical attempts to move from symptom-based categorization towards a more pathophysiological nosology, such as the symptom network theory \cite{borsboom2013network}. As such, AI (supported by access to big data) can help realign psychiatric diagnosis with biology.

Finally, machine-learning can offer a functional perspective on disease. Indeed, some would argue that attempting to isolate clinically-relevant neurobiological mechanisms from clusters of symptoms is unrealistic as the interaction between trauma, neurobiological alterations and symptom experience is neither unidirectional nor direct \cite{hinton2013local}. Instead, a diverse set of bodily, cognitive, social and cultural influences mediate these interactions at different time scales to maintain the clinical suffering. AI and big data technologies could help us monitor patients' ability to perform in daily activities and enjoy daily life, and model the individual factors relevant to each person's quality of life. As such it may no longer be necessary to categorize patients in a strict sense and the concept of psychiatric diagnosis could become obsolete. Instead, an understanding of each patient's genetic vulnerabilities, neurobiological states, clinical presentation, personal narratives and life trajectories and their interactions with others, as captured by an integrative AI model, would help clinicians target treatment and interventions without the need to label the patient with a diagnosis. This would create a new nosological system based on functionality and quality of life, which would more closely match patient needs and concerns \cite{robertson2009neurodiversity}.

\subsection{Prognosis}

One of the most complex steps in mental healthcare concerns prognosis, the understanding of which is important both on an individual patient level and from a public health policy standpoint. A prognosis is a prediction about the likely outcome of a patient's current disease (i.e. risk/chance of recovery, death, etc.). Prognosis for mental illness is often fairly unpredictable due to the following:
\begin{enumerate}
    \item Our perfunctory understanding of disease pathophysiology.
    \item Lack of well-documented and standardized longitudinal data on patients.
    \item Insufficient follow-up and dosage modulation from physicians.
\end{enumerate}


During recovery from mental illness, symptom improvement is not directly proportional to time elapsed during treatment (i.e. it is non-linear). However, by knowing the likely course a disease will take early on in the treatment process, including the likelihood of relapses and recurrences, we can better know which treatment to provide. For example, if we can predict that a given patient is likely to experience a relapse of the condition within a given time frame, it may be worth administering prophylactic treatment or scheduling more regular follow-ups with the clinician. Knowledge on the most likely course of a disease can also increase the efficiency of resource allocation and delivery of mental healthcare services. In particular, a patient's clinical team can better structure a long-term, integrative, and multifaceted care plan. Using AI, we can create personalized preventative mental health treatment based on these predictions. 

AI in mental illness prognosis can not only have a drastic impact on individuals, but also on society as a whole. Many countries, especially low-income and middle-income countries classify addressing mental health as low priority \cite{prince2007no}. Developing countries tend to prioritise the control of infectious diseases and reproductive health, which makes sense due to their urgency. That being said, in order to understand the public health impact of mental illness, we must consider that most mental disorders are diagnosed at a young age, where 75\% have an onset below the age of 24 \cite{prince2007no}. This is a vital point in an individual's life, both academically and socially, as this is the stage when these young adults begin their career, develop romantic relationships and life-long friendships. However, mental illness can have a drastic impact on the these seemingly normal steps in social development \cite{mezulis2014affective}. Additionally, several studies have correlated poor mental health with lower educational achievement \cite{patel2007mental}. Currently, a poor understanding of prognosis in psychiatry means that it may take several years for a patient to be properly treated \cite{patel2007mental}. 
On a population level, this leads to large groups of people missing the opportunity to pursue higher education and develop the appropriate social skills to become impactful members of society. In developing countries, this impact may be even higher as the future of the country relies on the development of an educated population who can contribute to the country's evolution. Hence, it would be worthwhile for developing countries to invest in the implementation of these AI-based technologies to help clinicians understand prognosis, thereby allowing patients to be properly treated at a younger age, and continue their pursuit of academic, professional and social success. A non-equitable implementation of AI-based care globally would further perpetuate the cycle of inequality facing persons in the developing world.

Furthermore, the high comorbidity rate between mental illness and other diseases indicates that improved prognosis in mental health could have a drastic impact even beyond the realm of psychiatry, also impacting diseases including cardiovascular illness and diabetes. Again, it is important for us to consider the complex relationships between mental illness and other diseases. It is possible that shifting a country's public health priority towards one that includes mental health may have a wider impact on other health conditions which are considered a priority. In terms of cost, a 2007 study discusses the economic impact of a phase-specific intervention program built for teens affected by psychotic disorders \cite{patel2007mental}. Importantly, the researchers identified that this method proved to be more cost-effective as costs shifted from in-patient services to community care. Employing AI to improve prognosis in mental health could allow for the development of improved, phase-specific interventions in depression as well. 

The use of AI technologies offers potential to gain greater insight on disease progression, potentially having a significant impact on healthcare delivery as a whole. The implementation of AI systems can help to shift the delivery of mental healthcare from a reactive response to a proactive response.

\subsection{Treatment Selection}

While determining prognosis can be helpful, it is often claimed that a prediction model is only as good as the system which uses it \cite{rosenfeld2018predicting}. In other words, instead of simply determining which patients will or will not improve- we should be evaluated based on the assignment of patients to the treatments that are most likely to be effective for them. This is often referred to as personalized, or precision, medicine.
 
Let us examine the treatment selection problem more closely. In mental health, many kinds of treatments exist- a range of pharmacotherapies, a range of psychotherapies, and neuromodulation techniques such as repetitive transcranial magnetic stimulation and electroconvulsive therapy. In addition \bsq{lifestyle} interventions, such as exercise, mindfulness, and meditation, have also been found to be effective for certain milder disorders or as adjuncts to medication or psychotherapy. What is striking is that most of the non-lifestyle interventions have been found to be roughly equally effective, despite disparate mechanisms and routes of administration \cite{bares2009low} (with the exception of electroconvulsive therapy, which for many conditions has superior efficacy but which is far more resource intensive than most other treatment options \cite{sackeim2017modern}. In addition, we must face the reality of resource restriction and the need to match patients to the right treatment intensity. This is exemplified by the \bsq{stepped care} approach \cite{bower2005stepped}, where patients are given access to the level of intervention they require and \bsq{stepped up} to more intense (and costly) services as needed. For example, in the UK's adult Improving Access to Psychological Therapies (IAPT) program \cite{clark2011implementing}, patients are streamed towards \bsq{low intensity} treatment (online resources and infrequent visits) or \bsq{high intensity} (weekly visits with a therapist). Patients are either streamed in to low intensity treatment first and are then moved up to high intensity treatment if the low intensity treatment fails, or they can be streamed directly into high intensity services based on diagnosis or symptom severity. Finally, because of the way mental health services are organized, it is often necessary to decide if a patient should be put on the waitlist for psychotherapy, should be started on medication, or should pursue medication and therapy concurrently.
 
In addition, when one is training a machine-learning model aimed at improving treatment selection, one must decide the \bsq{success} criteria. One could aim for the greatest amount of symptom reduction, though this may not always correlate with function or fully represent patient goals. One might aim for a reduction in suicide, though very large datasets may be needed in order to reliably detect such an effect given the low incidence of suicide. One might try and optimize cost effectiveness, though this might lead to slightly worse outcomes for many patients. Finally, when pursuing treatment selection it must be understood whether one is building a general model that can be applied to any patient population with a similar socio-demographic profile to the training set, or if one is optimizing a model which will help make treatment decisions within a specific healthcare system or institution. As such, it is clear that the treatment selection problem can be carved in many different ways, and important progress is now starting to be made on several different fronts and using a number of approaches. This progress, which we will now discuss, increases the hope that the right kind of big data, combined with a clear understanding of the treatment selection problem at hand and algorithms that are appropriate for solving it, can help significantly improve mental healthcare, reducing the time it takes for patients to find a treatment that works for them while reducing systemic costs by avoiding failed treatment courses.
 
Recent work has investigated several different perspectives of treatment selection prediction in depression. The Leeds Risk Index (LRI) identifies pre-treatment variables to predict treatment outcome and allows patients to be stratified into groups of low, moderate, or high risk for poor treatment response based on these LRI scores \cite{delgadillo2016different}. The baseline variables used to predict depression outcomes contain demographic and clinical information, including measures of age, employment status, disability status, and intellectual functioning. This type of risk index is proposed to have clinical relevance in directing patients towards treatment options of different intensity levels based on the predicted advantage for patients grouped by LRI scores. For example, the authors describe the value of this index in psychiatric treatment systems with discrete steps of treatment intensity such as those within the IAPT program; i.e. lower intensity treatment options such as providing support and teaching strategies based on Cognitive Behavioral Therapy (CBT), and higher intensity treatment options such as depression counselling or CBT sessions. The work suggests that low intensity treatment interventions may be the most cost-efficient approach for depressed patients with low LRI scores, while depression cases with high LRI scores may benefit more from high intensity treatment options and should avoid lower intensity treatment alternatives due to predicted higher dropout rates. This predictive insight has the potential to improve patient care at an individual level by assisting physicians in guiding patients towards the most effective treatment option, and improve patient care at a population level by distributing patients between differing treatment intensities based on predicted response patterns. This predictive approach to treatment selection has the potential to improve the efficiency of mental health treatment systems; in this case, by providing informed predictions to help patients and physicians navigate systems with discrete levels of treatment intensity. 

Neuroimaging data can also be used for differential treatment prediction. For example,  positron emission tomography (PET) imaging can be used to measure pre-treatment brain glucose metabolism in patients with depression receiving either escitalopram or CBT as treatment to predict response to treatment \cite{mcgrath2013toward}. The treatment-specific neuroimaging biomarker described by McGrath et al. suggests that distinct, observable, physiological differences can be measured using existing neuroimaging techniques to predict patient outcomes to  different treatments. On the other hand, the LRI suggests that stratification of patients with depression based on baseline demographic and clinical variables can predict the most beneficial treatment intensity level in a stepwise treatment system. These approaches, rather than competing, may provide complementary information regarding both the type of treatments and the intensity of the treatments that are most effective in individual cases.

In related work, \cite{derubeis2014personalized} created the Personalized Advantage Index (PAI). The PAI can be used to  identify both the most effective treatment as well as the magnitude of this benefit for an individual patient. The PAI identifies individuals who would benefit differentially between different treatments. Out of 154 patients in their study, 60\% of the sample displayed a clinically meaningful PAI score for one treatment compared to the other, meaning these patients are predicted to respond better to one of the two treatments. This PAI score was calculated by predicting symptom severity after treatment for each patient for paroxetine and CBT separately, then comparing the two estimates to determine the more beneficial treatment option. Clinically, identifying which patients would benefit differentially between treatments would allow for the prescription of the more effective treatment option for these individuals, and also allow patients without any discernible advantage for a specific treatment to select treatments based on personal patient values and potentially choose more cost efficient options. An important distinction was made by the authors of this work between prognostic variables, which predict non-specific treatment outcome, and prescriptive variables, which predict differential treatment outcomes. 

Overall, the above three approaches to differential treatment selection provide intriguing insight into the potential power of statistical analysis and modelling of data to more effectively guide treatment selection in the context of depression. From these different methodologies, it is clear that the future holds many exciting paths to improve treatment selection for mental healthcare upon the foundations of big data and rapidly advancing analytical and predictive tools.  

A recent successful application of machine-learning for treatment selection is \cite{benrimoh_fratila_israel_perlman_mirchi_desai_rosenfeld_knappe_behrmann_rollins_etal._2018}. The system, known as \textit{Aifred}, offers a neural network model that allows for differential prediction between the four different antidepressant drug categories. The model is capable of determining the overall likelihood of remission given each drug category. Using an extensive evaluation protocol, the system is found to provide a significant advantage over random drug allocation. The major contribution of this model is the extension of a differential benefit treatment selection process to \textit{more than two treatments}, which is key because of the large number of treatments available. A significant limitation of this analysis was the class imbalance in the training data- there were far more patients being treated with \bsq{citalopram}, than with any other drug. As such, we expect analyses of more balanced datasets to yield larger differential treatment prediction effect sizes. 

\subsection{Treatment Delivery}

Psychotherapy is an ancient form of healthcare that persists as an effective treatment option for a variety of mental illnesses, particularly affect disorders. However, psychotherapy delivery is hampered by two human-derived issues: 
\begin{enumerate}
    \item Limited access to appropriately qualified healthcare professionals due to cost and other logistical issues.
    \item  Variability in the quality of care.
\end{enumerate}

These limitations are now being addressed with machine-learning-based natural language processing and big data techniques.

A prominent example is \textit{Woebot}\footnote{\url{https://woebot.io/}} (Woebot) \cite{fitzpatrick_darcy_vierhile_2017}, a commercially-available psychotherapist, primarily powered by AI, with demonstrated short-term efficacy in reducing PHQ-9  scores amongst college students with self-identified symptoms of depression and anxiety. Users interact with Woebot via an instant messaging app, and these conversations are reviewed (typically at a later time) by a trained psychologist. Woebot's natural language is modelled after social discourse, and its response function decision tree is trained in CBT using three clinical sources \cite{burns_2007, burns_2009,towery_2016}. Six key process-oriented treatment features are prioritised: empathy, personalisation, goal-setting, accountability, motivation and reflection. Interestingly, some Woebot users who participated in the RCT by \cite{fitzpatrick_darcy_vierhile_2017} reported feeling a \say{real person concern} as a \say{most favoured feature} during intervention review, indicating that natural language is approaching the level of sophistication needed for elements of psychotherapy. Not all post-RCT reviews were positive of course, with complaints that Woebot \say{got a little repetitive} and that the conversations were inflexible and unnatural.

Another example is \textit{reSET}\footnote{\url{https://peartherapeutics.com/}} (Pear Therapeutics Inc.). reSET is a mobile app adjunct therapy for substance use disorder for patients abusing alcohol, cocaine, marijuana or stimulants. While reSET does not appear to directly utilise big data or machine-learning for therapy delivery, it does collect the types of patient data that could be leveraged to improve treatment prognosis \cite{bradway_joakimsen_grøttland_årsand_2016}. The more important aspect of reSET in the context of this chapter however, is in becoming the first FDA-approved digital therapeutic \cite{kennedy2018pear}.  

In recent years, mindfulness meditation \cite{davidson2003alterations} has surfaced as popular resilience training technique and there are hundreds of readily-available wellness apps claiming to be able to train users in mindfulness meditation. Unfortunately, the explosion of mindfulness apps in recent years has not been matched by a parallel explosion in RCTs examining mindfulness app efficacy, and it is still debated if these products can deliver tangible benefit, given mixed early results. To the authors' knowledge, to date only a handful of RCTs using active controls have interrogated commercially available mindfulness apps, most recently \cite{noone2018improvements}. These RCTs show a different outcome. In an examination of the app \textit{Headspace}\footnote{\url{https://www.headspace.com/}} (Headspace) as a 6-week intervention in undergraduate students, no benefits on wellbeing, affect, cognitive function or mindfulness abilities were revealed, either through within-subject pre-post analysis or in comparison to the active control. In contrast, in an examination of the app \textit{Wildflowers}\footnote{\url{http://www.midigitaltherapeutics.com/}} (Mobio Interactive Inc., of which BS and MT are a part) as a 3-week intervention in undergraduate students, benefits to well-being and stress-resilience were revealed in both within-subject analysis as well as in comparison to the active control. 

Interestingly, a unique aspect of Wildflowers and its sister products developed by Mobio Interactive is the use of big data and machine-learning. For example, Wildflowers leverages computer vision to extract heart rate variability through photoplethsymographic imaging \cite{chwyl2016sapphire,chwyl2016time} of user selfie videos. Since heart rate variability is negatively correlated with cognitive stress \cite{thayer2012meta}, this technology has the potential to objectively quantify stress. According to the Mobio Interactive website, over 100,000 pairs of selfie videos and self-assessments of mood and stress from users throughout the world are currently being leveraged to train deep neural networks that both objectively and remotely predict stress changes in the end user, and then personalise psychotherapy accordingly. It remains to be determined if the use of big data and machine-learning in this context contributes to clinical efficacy, but given how effectively big data and machine-learning have been applied in the various contexts described throughout this chapter, it seems likely that such practices will ultimately give rise to more efficacious digital therapeutics for the patient.

\subsubsection{Special Opportunities}

\textit{Real World Validation.} With almost as many cell-phone subscriptions as there are humans and penetration rates in developing countries averaging 90\%, the potential to use mobile devices for health-related data generation is unsurpassed by any other previously available data collection method in history. With these real world data comes the potential for verifying \textit{real world efficacy}, i.e obtaining \textit{real world validation}. Real world validation offers an unprecedented opportunity for transparency in healthcare. Digital therapeutics like the ones mentioned above and others may one day soon stream live anonymous and objective data on stress and wellbeing, continually monitoring the real world efficacy of each product in realtime. 

\textit{Big Data Loop. } Without question, the greatest promise of combining big data and machine-learning with digital therapeutics is the generation of a \say{Big Data Loop} that enables seamless feedback circuitry to continuously refine therapy and prognosis in realtime as data stream in for analyses. In this context, the real-time collection and analysis of real world data from digital therapeutics plays a central role in redefining the relationship between large numbers of patients and health care providers – in some cases providing the first, or even the only, point of contact between health care systems and individuals challenged with mental health conditions.

\subsubsection{Specific Challenges}

\textit{Public Acceptance and Adoption.} Pharmaceuticals have dominated healthcare for about a century, and became a first-line treatment option for mental illness beginning in the 1970's. At present, it is well accepted that small molecule pharmaceuticals have definitive biological effect, and often of a net-positive nature. The same patient confidence cannot be stated for the digital forms of therapy that are guided by AI or required to gather critical patient data. This lingering skepticism is likely to slow adoption of digital products until sufficient real world efficacy permeates the public consciousness.

\textit{Differentiation.} Given that much of the public already demonstrates clear confusion between approved clinical practices and medical hoaxes (e.g. homeopathy), it should be expected that at least as much of the public will find it difficult to differentiate between digital therapeutics that are backed by scientific evidence from the large number that are not, or that have even failed RCTs \cite{noone2018improvements}. The generation of a consumer friendly and tightly controlled cross-border e-commerce site (i.e. a \say{medical app store}) for public-facing digital therapeutics may be one viable solution.

While these are still early (albeit exciting) days, AI-powered and/or big data collecting psychotherapeutic interventions like Woebot, reSET and Wildflowers are likely to have a massive positive impact on treating mental health, globally. People in all countries and from all walks of life use their mobile devices every day. These interfaces may soon deliver affordable and effective mental health care.

\subsection{Monitoring}

Continuous monitoring of patients plays a critical role in mental healthcare, for many reasons: Firstly, the mental health clinician (psychiatrist, therapist, etc.) receives a very partial view of the full condition of the  patient. Typically, non-hospitalized patients do not see their clinician  on a very frequent basis and therefore the treatment is based only on the report of patient at the appointments. Secondly,  symptoms can change significantly over a short period and thus necessitate more immediate intervention. In addition, mental illness is often episodic, in particular for patients suffering from depression or schizophrenia in which recurrences or relapses are common. Such patients would benefit from their  clinician receiving regular updates of their symptoms.
 
A possible na\"ive approach for performing monitoring would be to manually contact the patient frequently in order to receive reports of mood and symptoms. However, this approach is not-efficient since it requires great effort from both the patient and  the clinician. In addition, self reporting suffers from a lack of accuracy. Therefore, in the past two decades, many different methods and applications have been developed  for automatic monitoring of mental health. 

Advances in technology enable the effective \textit{collection} of monitored data through several forms and means. Specifically, smartphones are an extremely valuable tool for monitoring mentally ill patients, since they have become very common among the overall population in Western countries and specifically among mental health patients and are carried by the patients throughout most of the day  \cite{firth2015mobile}. In addition, smartphones are continuously improving in memory storage and processors capabilities for recording and processing information. Other forms of collection of monitores data can be done through computers and designated wearable devices. Automatic monitored data can be classified in two types: 
\begin{enumerate}
    \item \textit{Subjective data}, such as patient's  self report of mood and symptoms in response to a mobile application's daily inquiry.
    \item \textit{Objective data}, such as behavioral data (e.g., activity, phone usage), physiological data (e.g. heart rate, body temperature) or environmental information (e.g. location, outdoor exposure). 
\end{enumerate}

Each of these  types of data have advantages and limitations: Objective measures are naturally more accurate and can capture a large amount of information without interrupting the patient's daily routine. However, these measures are limited since in many mental-illnesses the diagnosis depends to a great extent on the patient's description of feeling and mood. Subjective measures, on the other hand, are often affected by the context of their assessment and biased by the mood of the patient. 

The research in the field of automatic monitoring of mental health has focused mainly on the feasibility of collection of data and on finding associations between the monitored data and the patient's mood and symptoms.   However, almost none of these studies have attempted to create an application which would utilize the data to assist in real time decision making regarding treatments.  Some studies have introduced systems which perform interventions in the treatment, however the intervention are relatively simple, and  to  the best of our knowledge, no application includes advanced tools of AI, such as autonomous agents \cite{russell2016artificial}.

In the following, we will describe the existing research and the applications of monitoring mental health, and we will layout future directions for research.

\subsubsection{Symptom Monitoring}
 
Many of the studies in the field of symptom monitoring have explored the association of clinical states, or transitions in mood states, with monitored data that can be automatically and passively monitored.
 
Some studies have found that physiological  measures can predict clinical states. For example, Lantana et al. \cite{lanata2015complexity} introduce \textit{PSYCHE}, a personalized wearable monitoring system, designed to improve the management of patients suffering from mental disorders and specifically Bipolar Disorder. PSYCHE is a t-shirt with embedded sensors, which monitor the Heart-Rate Variability.  The authors demonstrate that PSYCHE is successful in assessing transitions from pathological mood states.
 
Other studies have shown that the phone usage patterns of patients can indicate patients' mood state \cite{burns2011harnessing, grunerbl2012towards, muaremi2014assessing}. For example, Gr\"{u}nerbl et al \cite{grunerbl2012towards} found that the duration and frequency of phone calls increase for individuals with mild depressive disorders compared to individuals with severe depression or a normal mood state. In addition, physical activity has been shown to be linked to affective states \cite{abdullah2016automatic, beiwinkel2016using, faurholt2012differences, dickerson2011empath, grunerbl2012towards}. In general, individuals with affective disorders tend to have lower activity energy and acceleration compared to healthy individuals \cite{faurholt2012differences}. Furthermore, studies have found a relationship between the emotional state and the following categories of data monitored by smartphones: voice features  \cite{abdullah2016automatic, dickerson2011empath, muaremi2014assessing}, light exposure \cite{burns2011harnessing, abdullah2016automatic} and location changes \cite{burns2011harnessing, abdullah2016automatic, gruenerbl2014using, grunerbl2012towards}.
 
Monitored data can also be useful for predicting and preventing relapse. Relapse prevention is specifically an important issue among patients diagnosed, hospitalized, and treated for schizophrenia, since up to 40\% of those discharged may relapse within a year (even with appropriate treatment) \cite{barnett2018relapse}. In \cite{barnett2018relapse}, Barnett et al. identified statistically significant anomalies in patient behavior, as measured through smartphone use, in the days prior to a relapse.
 
A major part of the research in the field has focused on the feasibility of using monitoring devices and patients' adherence to user guidelines. Many studies which evaluated smartphone-based monitoring have reported patient compliance as a limitation \cite{abdullah2016automatic, beiwinkel2016using, gruenerbl2014using, dang2016accompanying}, as some patients did not carry their phones with them all of the time or occasionally turn their phones off. In one study, patients noted that they would be interested in using the monitoring smartphone more regularly if transparency concerning the recorded data a was guaranteed \cite{dang2016accompanying}.
 
Only a small portion of the studies in this field have investigated the clinical outcomes of automated monitoring for mental health. In \cite{saunders2017experiences}, Saunders et al. conducted a longitudinal study in which individuals with bipolar disorder monitored their mood daily using a smartphone application for twelve weeks. In a follow-up interview, half of the participants noted that they had improved their mood since they were able to better recognize their feelings, and half of the participants also reported a change in behavior (e.g. increased exercise levels).  In \cite{wu2018comparative}, Wu et al. evaluated in a six-month study an automated telephone assessment systems which monitored patients with depression and type 2 diabetes by calling them regularly. The call contents were individually determined through an algorithm that scanned patient medical records and call histories to determine applicable questions. The system also alerted emergency responders to contact immediately patients who exhibited suicidal ideation. The automated system significantly increased  both associated depression remission and patient satisfaction compared to the control group.
 
\subsubsection{Monitoring Compliance to Treatment}
 
Another form of monitoring which has been investigated in research is the monitoring of patient compliance with treatment. Non-adherence to psychotropic medication is a significant issue in mental health treatment. For example, in Bipolar Disorders, estimates of non-adherence range between 20\% and 60\%  treatment, with non-adherence often leading to negative outcomes \cite{levin2015use}.
 
Some studies have tested the effect of medication-adherence tele-monitoring systems which record the date and time of the medication bottle openings. For example, Frangou et al \cite{frangou2005telemonitoring} evaluated a system that included an electronic dispenser that fits on the medicine bottle cap and records the date of each bottle opening. The data was automatically transmitted online to clinicians who received alerts if adherence dropped below 50\%. They tested the systems efficacy among individual with schizophrenia and found that it significantly improved medication adherence and improved the psychotic symptoms in comparison to the control group. 
 
Bickmore et al \cite{bickmore2010maintaining} tested an automated system, including of an animated agent, which conducts simulated conversations in order to promote medication adherence in individuals with schizophrenia by establishing an emotional relationship with the patient and providing consistent social support. They conducted a one-month pilot study, in which individuals with schizophrenia were instructed to have daily interactions with the agent. During the interaction, the agent inquired about medication adherence and provided tips and suggestions for solving adherence problems. They found that the agent was successful in increasing the rate of adherence among the participants.
 
An additional problem (regarding compliance to treatment) associated with poor therapeutic outcomes is non-attendance of psychotherapy sessions. In  \cite{bruehlman2017mobile}, Bruehlman-Senecal et al. found that automated mood-monitoring text messages can be used as a predictor of  psychotherapy attendance.

 


\subsection{Ethical Considerations}

Psychiatry faces unique challenges in addition to those common to all health care disciplines. While we have discussed how AI can offer promising solutions to those challenges, from diagnosis to treatment selection, such implementations create important ethical considerations that need to be addressed if the benefits are to be realized without causing undue harm. This section focuses on ethical issues specific to the use of AI and big-data approaches in mental health. 

AI algorithms are only as good as the data they are trained on. A widely-discussed concern with the use of AI in high-stakes applications like healthcare involves the quality and quantity of data needed and the possible replication or amplification of biases present in the data \cite{mittelstadt2016ethics}. This issue is particularly concerning in the mental health sector. First, mental health conditions will probably require more and more diverse data than physical conditions to accurately picture their complexity. Indeed, mental disorders are multisystem disorders- affecting mood, perception, cognition and volition- and are caused by a complex interaction of more proximal biological causes and more distal environmental causes \cite[Chapter~2]{schaffner2008etiological}. The absence of any accurate etiological or pathophysiological models prevent us from pre-selecting relevant features as we do not know how they interact. While it is hoped that AI will help us uncover such interactions, extensive data from every levels or dimensions is deemed necessary to avoid biases. Second, many factors also threaten the quality of mental health data: patients with mental disorders are more subject to treatment non-compliance and high drop-out rates in clinical studies  \cite{chen1991noncompliance}. Also, the prominent stigmatisation around mental health can threaten reliability of patients' reports. Moreover, the lack of mental illness biomarkers renders mental health data less quantifiable in general compared to other health conditions. Psychiatric terminology often involves concepts that are subjective and can be interpreted in many ways, making comparing stories of individual patients sometimes more difficult. All of this increases the risk that the algorithm makes erroneous advice in a mental health context. Potential social or governmental action to improve mental health services in order to ensure optimal data accessibility and quality may help avoid any added discrimination (i.e. via biased algorithm results, or lack of access to useful algorithms because of a paucity of quality data) in an already heavily stigmatized population. 

Another well-acknowledged issue in AI relates to the interpretability or transparency of the algorithm \cite{mittelstadt2016ethics}, that is the ability to understand the step-by-step path taken by the algorithm to arrive at its conclusion. Transparency seems necessary in order to avoid conflict of interest and potential malicious use of Clinical Decision-Support Systems (CDSS) that are aimed at helping clinicians and patients make decisions. For example, a treatment selection CDSS could be programmed to output certain drugs more often in order to generate higher profits for their designers instead of prioritizing clinical outcomes \cite{char2018implementing}. This is particularly relevant in mental health care where many lines of treatment exist, but there is no systematic procedure for selecting between them (as discussed before). Moreover, as comprehensibility of an algorithmic decision cannot be made without full knowledge of all features that are inputted, transparency runs against other ethical ideals, like privacy of data subjects. Medical confidentiality is a prime principle of medical care, of special importance in the psychiatric setting considering the high level of stigma around mental health. If any third party is in charge of interpreting and explaining a CDSS output to the physician or patient, this could jeopardize medical confidentiality, as they would potentially have access to the patient's sensitive features. To prevent such concerns, efforts and resources should be directed to train the physician and patient on the AI tools and ensure their autonomy on CDSS usage and interpretability, or ensure that interpretability report generation by third parties occurs without these parties having access to identifying patient information. 

Third, a trusting patient-physician relationship represents a central component of mental health care. AI products will undoubtedly re-shape this relationship's dynamics, whether it is a CDSS used only by the physician, an interface to improve daily communication between the patient and their physician or an automated conversational agent used only by the patient \cite{fitzpatrick_darcy_vierhile_2017}. An evident related concern is the responsibility and liability associated with algorithmic decisions \cite{Bla}. It seems collectively understood that AI should not replace clinical judgement and that physician input should remain critical at every stage of the clinical process. While relying on physician judgement seems justified for now as algorithms are still limited and laden with potential biases \cite{friedman1996bias}, it will become less evident as this technology continues to develop and improves in accuracy and quality. Would a physician be liable if they disregard advice of a high-quality CDSS and this results in harm coming to the patient? There are also more subtle ways in which AI can be detrimental to the patient-physician relationship. The AI \say{narrative}, i.e. the ways of talking about AI and the terms used by the physician in their clinical encounter with the patient, will most certainly have dramatic effects on patient wellbeing. The placebo effect and physician and patient expectations have long been recognized as playing a significant role in treatment efficacy \cite{benson1975placebo}. This is particularly true in the field of psychiatry \cite{kirsch2014antidepressants} where we do not know exactly how most treatments work to relieve symptoms. Considering the hype around AI on the one hand, but the distorted views of its potential risks being presented to the public on the other hand, it will be imperative to regulate the place of AI in the patient-physician relationship and educate clinicians and patients on AI, big data, and its limitations to avoid deception or a blind compliance of the patient or clinician with AI recommendations. 

\section{Conclusions}\label{sec:conclusions}

As reviewed in this chapter, the significance of AI and big data in mental healthcare cannot be overstated. The use of these technologies will facilitate diagnosis, prognosis prediction, treatment selection and delivery, disease monitoring, and will optimize the allocation of healthcare resources, all of which be utilized to inform public health policy. Making use of technological innovations in the field of mental healthcare is primordial in the quest to tackle the current inefficiency of the system, especially considering the fact that the expected burden of mental illnesses is rising over time \cite{thyloth2016increasing}. Indeed, medical professionals are increasingly recognizing the importance of harnessing big data to rectify the dysfunction inherent in the current system, which is one of the reasons that clinicians are adopting the framework set forth by the National Institute of Mental Health (NIMH) called the Research Domain Criteria (RDoC) of mental health classification. The data-driven RDoC is \say{[...] an attempt to create a new kind of taxonomy for mental disorders by bringing the power of modern research approaches in genetics, neuroscience, and behavioral science to the problem of mental illness} \cite{press}. This quantitative system focused on biology, while still making use of symptom tracking and other subjective metrics, plays a new and important role by filling in pieces that have been missing from the complex puzzle of mental illness diagnosis and treatment.
 
While the introduction of big data into mental healthcare will bring about widespread social and economic benefits, it will also generate its own unique and unprecedented challenges. The most salient of these challenges will be to enforce an ethical development and equitable delivery of AI solutions. Additionally, any robust AI model must be built with comprehensive data to encompass all possible treatments types. Importantly, data from people of all races, ethnicities, and socioeconomic backgrounds must be used in model training to avoid bias.
 
Using big data and machine-learning to capture and quantify the heterogeneity within patient diagnosis and treatment response can elucidate the biological mechanisms underlying the diseases themselves. In other words, the data collected to feed an AI model will inevitably prompt research questions and hypotheses by highlighting particular variables that are salient in the model's decision-making processes. For example, if the AI finds that a series of immune markers were heavily weighted in comparing predictions of treatment outcome between two antidepressant medications, then this implies that one medication may be targeting some sort of immune dysregulation, adding weight to the hypothesized link between depression and the immune system. AI is therefore necessary in mental health research in order to disentangle the non-linear relationships between potential predictive factors and distill individual factors with robust predictive power. In short, data from basic research will feed the AI, and results from the AI will also feed basic research.  
 
While this is an exciting time for mental healthcare, a technological reform of such a scale must be implemented in a proactive, careful, and deliberate manner. One must remember that data points in a machine learning model are representative of real people who are suffering from real mental illnesses. Therein lies the true value of big data in mental health - bringing personalized treatment to a field of medicine in which it is so desperately lacking.

\section*{References}

\bibliography{bib,bib2,bib3}

\end{document}